\documentclass[11pt]{article}
\usepackage{amsfonts}
\usepackage{amssymb}
\usepackage{amsmath}
\usepackage{graphicx}
\usepackage{geometry}

\newtheorem{theorem}{Theorem}

\newtheorem{definition}{Definition}

\newtheorem{proposition}{Proposition}
\newtheorem{remark}{Remark}

\makeatletter
\def\@removefromreset#1#2{\let\@tempb\@elt
     \def\@tempa#1{@&#1}\expandafter\let\csname @*#1*\endcsname\@tempa
     \def\@elt##1{\expandafter\ifx\csname @*##1*\endcsname\@tempa\else
    \noexpand\@elt{##1}\fi}     \expandafter\edef\csname cl@#2\endcsname{\csname cl@#2\endcsname}     \let\@elt\@tempb
     \expandafter\let\csname @*#1*\endcsname\@undefined}

\@removefromreset{equation}{section}

\@removefromreset{theorem}{section}
\makeatother
\input{tcilatex}
\begin{document}

\title{Quantum analog of the original Bell inequality for two-qudit states
with perfect correlations/anticorrelations}
\author{Elena R. Loubenets$^{(1)}$ and Andrei Y. Khrennikov$^{(2)}$ \\
$^{(1)}$ Applied Mathematics Department, National Research University \\
Higher School of Economics, Moscow, 101000, Russia, elena.loubenets@hse.ru\\
$^{(2)}$ International Center for Mathematical Modeling, Linnaeus \\
University, 35195 Vaxjo, Sweden, Andrei.Khrennikov@lnu.se}
\maketitle

\begin{abstract}
For an even qudit dimension $d\geq 2,$ we introduce a class of two-qudit
states exhibiting perfect correlations/anticorrelations and prove\ via the
generalized Gell-Mann representation that, for each two-qudit state from
this class, the maximal violation of the original Bell inequality is bounded
from above by the value $3/2$ -- the upper bound attained on some two-qubit
states. We show that the two-qudit Greenberger--Horne-- Zeilinger (GHZ)
state with an arbitrary even $d\geq 2$ exhibits perfect
correlations/anticorrelations and belongs to the introduced two-qudit state
class. These new results are important steps towards proving in general the $%
3/2$ \emph{upper bound }on quantum violation of the original Bell
inequality. The latter would imply that similarly as the Tsirelson upper
bound $2\sqrt{2}$ specifies the quantum analog of the CHSH inequality for
all bipartite quantum states, the upper bound $\frac{3}{2}$ specifies the
quantum analog of the original Bell inequality for all bipartite quantum
states with perfect correlations/ anticorrelations. Possible consequences
for the experimental tests on violation of the original Bell inequality are
briefly discussed.
\end{abstract}

\section{Introduction}

The Clauser-Horne-Shimony-Holt (CHSH) inequality \cite{CHSH} on local
classical correlations was introduced in 1969 and in 11 years Tsirelson \cite%
{Ts1, Ts2} proved that, for any bipartite quantum state, possibly, infinite
dimensional, the maximal violation of the CHSH inequality cannot exceed $%
\sqrt{2}$ and that on some two-qubit states this upper bound is attained.

The original Bell inequality for local classical correlations was derived 
\cite{B1, B2} by Bell even earlier -- in 1964, however, to our knowledge,
until our recent article \cite{KL} on quantum violation of the original Bell
inequality by general two-qubit and two-qutrit states under spin
measurements, the maximal quantum violation of this inequality was
considered in the literature only for two-qubit Bell states and, mostly, for
the singlet -- Pitowsky \cite{Pit} and Khrennikov\&Basieva \cite{KB} proved
that, for the two-qubit singlet state, the maximal violation is equal to $%
\frac{3}{2}$ and stressed that this value is more than the Tsirelson \cite%
{Ts1, Ts2} upper bound $\sqrt{2}$ on the maximal violation of the CHSH
inequality.

There are probably three major reasons for disregarding the problem of
finding the maximal violation of the original Bell inequality for a general
bipartite quantum state with perfect correlations/anticorrelations.

First of all, only the two-qubit singlet state was known to satisfy the
condition on the perfect anticorrelation of outcomes whenever \emph{each }%
qubit spin observable is measured at both sites -- the Bell condition
sufficient for the derivation of the original Bell inequality in a local
hidden variable (LHV) frame.

Note that, for the derivation of the original Bell inequality in a local
hidden variable model, the Bell restriction $\mathrm{tr}[\rho \{B\otimes
B\}]=\pm 1$ on a bipartite quantum state $\rho $ on a Hilbert space $%
\mathcal{H}\otimes \mathcal{H}$ and a quantum observable $B$ on $\mathcal{H}$
is only sufficient but not necessary. In the LHV model, the validity of this
condition for only some observable $B$ with eigenvalues in $[-1,1]$ implies
the validity for a state $\rho $ of the original Bell inequality%
\begin{equation}
\left\vert \text{ }\mathrm{tr}[\rho \{A\otimes B\}]-\mathrm{tr}[\rho
\{A\otimes \widetilde{B}\}]\text{ }\right\vert \pm \mathrm{tr}[\rho
\{B\otimes \widetilde{B}\}]\leq 1  \label{01}
\end{equation}%
(in its perfect correlations (plus sign) form or perfect anticorrelations
(minus sign) form) for all observables $A,\widetilde{B}$ with eigenvalues in 
$[-1,1].$ Moreover, as proved in \cite{L1}, for any dimension of a Hilbert
space $\mathcal{H}\otimes \mathcal{H},$ there is the whole class of
bipartite quantum states that satisfy the perfect correlation form of
inequality (\ref{01}) for any three quantum observables $A,B,\widetilde{B}$
with eigenvalues in $[-1,1]$ but do not need to exhibit the perfect
correlation or anticorrelation of outcomes if some qubit observable is
measured at both sites (perfect correlations/anticorrelations, for short).

Secondly, as we have discussed this in Section 2 of \cite{KL}, the Tsirelson 
\cite{Ts1, Ts2} upper bound $2\sqrt{2}$ for the CHSH inequality implies the
upper bound $\left( 2\sqrt{2}-1\right) $ for the maximal value of the
left-hand side of inequality (\ref{01}) in any bipartite quantum state with
perfect correlations/anticorrelations. Therefore, specifically the latter
bound was considered to be the least one for violation of the original Bell
inequality in a quantum perfect correlation/anticorrelation case. However,
as proved in \cite{Pit, KB}, for the singlet, the maximal value of the left
hand side of inequality (\ref{01}) is equal to $\frac{3}{2}$ and this value $%
\sqrt{2}<\frac{3}{2}<2\sqrt{2}-1$.

Thirdly, for more than 40 years since derivation of the original Bell
inequality it was impossible to approach high fidelity in preparation of
two-qubit singlet states. Only recently experimenters approached very high
levels of fidelity. Therefore, the original Bell inequality was considered
as a theoretical statement without any possibility for its experimental
verification. The absence of possible experimental applications lowered the
interest of theoreticians to this inequality and the main theoretical
efforts were put into analysis of the CHSH inequality. We remark that even
nowadays the experimental testing of the original Bell inequality is a big
challenge, since one has to combine high levels of purity in preparation of
the two qubit singlet state and of detection efficiency (see \cite{KB} for
details).

Analyzing in \cite{KL} the maximal violation of the original Bell inequality
in a general two-qubit case, we introduced \emph{a necessary and sufficient
condition }for a symmetric two-qubit state to exhibit perfect
correlations/anticorrelations. We proved \cite{KL} that, for each symmetric
two-qubit state with perfect correlations/anticorrelations, the maximal
violation of the original Bell inequality under spin measurements is upper
bounded by the value $\frac{3}{2}$ and specified two-qubit states where this
upper bound is attained. We also considered \cite{KL} spin measurements on
symmetric two-qutrit states with perfect correlations/anticorrelations and
found that, in this case, the maximal violation of the original Bell
inequality is also upper bounded by $3/2$. \ 

Therefore, in Conclusions of \cite{KL}, we assumed that, similarly as the
Tsirelson upper bound $2\sqrt{2}$ specifies \cite{Ts1, Ts2} the quantum
analog of the CHSH inequality for all bipartite quantum states and all
quantum observables with eigenvalues in $[-1,1]$, the upper bound $\frac{3}{2%
}$ specifies the quantum analog 
\begin{equation}
\left\vert \text{ }\mathrm{tr}[\rho \{A\otimes B\}]-\mathrm{tr}[\rho
\{A\otimes \widetilde{B}\}]\text{ }\right\vert \pm \mathrm{tr}[\rho
\{B\otimes \widetilde{B}\}]\leq \frac{3}{2}  \label{02}
\end{equation}%
of the original Bell inequality for all bipartite quantum states with
perfect correlations/ anticorrelations: $\mathrm{tr}[\rho \{B\otimes
B\}]=\pm 1.$

In the present paper, for an arbitrary even $d\geq 2,$ we introduce a class
of symmetric two-qudit states exhibiting perfect
correlations/anticorrelations and show that, for all two-qudit states from
this class, the maximal violation of the original Bell inequality over
traceless qudit observables with eigenvalues $\pm 1$ is bounded by $3/2$
from above. We prove that, for any even $d\geq 2,$ the two-qudit
Greenberger--Horne--Zeilinger (GHZ)\emph{\ }state exhibits perfect
correlations/ anticorrelations and belongs to the introduced state class.

These new results are important steps towards proving that inequality (\ref%
{02}) constitutes the quantum analog of the original Bell inequality for all
bipartite quantum states with perfect correlations/anticorrelations.

The paper is organized as follows.

In Section 2, we recall the main issues on the derivation of the original
Bell inequality in a local hidden variable frame.

In Section 3, we specify shortly our previous results \cite{KL} on the
maximal violation of the original Bell inequality under spin measurements on
arbitrary two-qubit states and two-qutrit states with perfect
correlations/anticorrelations.

In Section 4, due to the properties of the generalized Gell-Mann
representation for traceless qudit observables proved in \cite{L3}, we
consider violation of the original Bell inequality in a general two-qudit
case ($d\geq 2).$ We show (Proposition 1) that, for an arbitrary even $d\geq
2,$ there exists the whole class of states exhibiting perfect
correlations/anticorrelations and that the GHZ state belongs (Proposition 2)
to this state class. We further prove (Theorem 1) that, for each state from
this class, the maximal violation of the original Bell inequality is bounded
from above by the value $3/2.$

In Section 5, we summarize the main results of the present paper.

\section{Preliminaries: the original Bell inequality}

Consider a bipartite correlation scenario\footnote{%
On the general framework for the description of a multipartite correlation
scenario, see \cite{L2}.} where two parties (say, Alice and Bob) perform
measurements, indexed by $a_{i},b_{k},$ $i,k=1,2,$ and with outcomes $%
\lambda _{a},\lambda _{b}\in \lbrack -1,1]$ at Alice and Bob sites,
respectively. This correlation scenario is described by four joint
measurements $(a_{i},b_{k}),$ $i,k=1,2$, and, for a measurement $%
(a_{i},b_{k}),$ notation $P_{(a_{i},b_{k})}(\lambda _{a},\lambda _{b})$
means the joint probability of the event that Alice observes an outcome $%
\lambda _{a}$ and Bob -- an outcome $\lambda _{b}$, and the expectation
value (average) of the product of their outcomes is given by 
\begin{equation}
\langle \lambda _{a_{i}}\lambda _{b_{k}}\rangle :=\sum_{\lambda _{a},\lambda
_{b}\in \lbrack -1,1]}\lambda _{a}\lambda _{b}P_{(a_{i},b_{k})}(\lambda
_{a},\lambda _{b})  \label{1}
\end{equation}

For the above correlation scenario, consider the value of the Bell \cite{B1,
B2}) combination of the product expectations 
\begin{equation}
\left\vert \text{ }\langle \lambda _{a_{1}}\lambda _{b_{1}}\rangle -\langle
\lambda _{a_{1}}\lambda _{b_{2}}\rangle \right\vert \pm \langle \lambda
_{a_{2}}\lambda _{b_{2}}\rangle  \label{2}
\end{equation}%
under the Bell condition \cite{B1, B2} on perfect
correlations/anticorrelations:%
\begin{equation}
\langle \lambda _{a_{2}}\lambda _{b_{1}}\rangle =\pm 1.  \label{3}
\end{equation}%
In view of (\ref{1}), this condition takes the form:%
\begin{equation}
\langle \lambda _{a_{2}}\lambda _{b_{1}}\rangle =\pm 1\text{ \ \ }%
\Leftrightarrow \text{ \ \ }\sum_{\lambda _{a},\lambda _{b}\in \lbrack
-1,1]}(\lambda _{a}\lambda _{b}\mp 1)P_{(a_{i},b_{k})}(\lambda _{a},\lambda
_{b})=0.  \label{4}
\end{equation}%
Since $\lambda \in \lbrack -1,1],$ this implies that the Bell condition (\ref%
{3}) is fulfilled under measurement $(a_{2},b_{1})$ if and only if, for
outcomes $\lambda _{a},$ $\lambda _{b}$, with the product $\lambda
_{a}\lambda _{b}\neq \pm 1,$ the corresponding joint probability $%
P_{(a_{2},b_{1})}(\lambda _{a},\lambda _{b})=0$.

Let the probabilistic description of this correlation scenario admits a
local hidden variable (LHV)\ model in the sense that all joint probabilities 
$P_{(a_{i},b_{k})},$ $i,k=1,2,$ admit the representation 
\begin{eqnarray}
P_{(a_{i},b_{k})}(\lambda _{a},\lambda _{b}) &=&\int\limits_{\Omega
}P_{a_{i}}(\lambda _{a}|\omega )P_{b_{k}}(\lambda _{b}|\omega )\text{ }\nu (%
\mathrm{d}\omega ),\text{ \ \ }  \label{6} \\
\lambda _{a},\lambda _{b} &\in &[-1,1],\text{ \ }i,k=1,2\text{,}  \notag
\end{eqnarray}%
via a single probability distribution $\nu $ of some variables $\omega \in
\Omega $ and conditional probability distributions $P_{a_{i}}(\mathrm{\cdot }
$ $|\omega ),$ $P_{b_{k}}(\mathrm{\cdot }$ $|\omega )$ of outcomes at
Alice's and Bob's sites, each distribution depending\footnote{%
Independence of distributions $P_{a_{i}}(\mathrm{\cdot }$ $|\omega ),$ $%
P_{b_{k}}(\mathrm{\cdot }|\omega )$ on setting of other measurements is
referred to as Bell locality, see \cite{L4} for details.} only on a setting
of the corresponding measurement at the corresponding site.

Then, under condition (\ref{3}) on perfect correlations/anticorrelations,
the Bell expression (\ref{2}) for product expectations satisfies the
original Bell inequality \cite{B1}: 
\begin{equation}
\text{ }\left( \left\vert \text{ }\langle \lambda _{a_{1}}\lambda
_{b_{1}}\rangle -\langle \lambda _{a_{1}}\lambda _{b_{2}}\rangle \right\vert
\pm \langle \lambda _{a_{2}}\lambda _{b_{2}}\rangle \right) {\LARGE |}_{%
\text{perfect}}\leq 1  \label{7}
\end{equation}%
in its perfect correlation (plus sign) or perfect anticorrelation (minus
sign) forms. For the proof of inequality (\ref{7}) in an LHV model for
arbitrary $\lambda _{a},\lambda _{b}\in \lbrack -1,1]$ and a more general
(than (\ref{3})) sufficient condition for its validity, see \cite{L1}.

\section{Quantum violation by two-qubit and two-qutrit states}

Let, under a bipartite correlation scenario, two parties perform
measurements on a two-qudit state $\rho _{d\times d}$ on $\mathbb{C}%
^{d}\otimes \mathbb{C}^{d},$ $d\geq 2,$ and measure traceless qudit
observables $A_{1},$ $A_{2}=B_{1},B_{2}$ with eigenvalues in $[-1,1]$.

In a quantum case, the product expectations (\ref{1}) take the form 
\begin{equation}
\langle \lambda _{a_{i}}\lambda _{b_{k}}\rangle =\mathrm{tr}[\rho _{d\times
d}\{A_{i}\otimes B_{k}\}],  \label{8}
\end{equation}%
and expression (\ref{2}) (the left hand-side of the original Bell inequality
(\ref{7})) and the Bell condition (\ref{3}) reduce to 
\begin{equation}
\mathcal{B}_{\rho _{d\times d}}^{ob}(A,B^{(\pm )},\widetilde{B})=\left\vert 
\text{ }\mathrm{tr}[\rho _{d\times d}\{A\otimes B^{(\pm )}\}]-\mathrm{tr}%
[\rho _{d\times d}\{A\otimes \widetilde{B}\}]\text{ }\right\vert \pm \mathrm{%
tr}[\rho _{d\times d}\{B^{(\pm )}\otimes \widetilde{B}\}],  \label{9}
\end{equation}%
and 
\begin{equation}
\mathrm{tr}[\rho _{d\times d}\{B^{(\pm )}\otimes B^{(\pm )}\}]=\pm 1,
\label{10}
\end{equation}%
respectively, where, for short, we change notations $A_{1}\rightarrow A,$ $%
B_{1}\rightarrow B^{(\pm )},$ $B_{2}\rightarrow \widetilde{B}$.

\begin{remark}
Similarly as we have discussed this above in Section 2 (see Eq. (\ref{4}),
let us analyze, for what a quantum observable $B^{(\pm )}$ with eigenvalues
in $[-1,1],$ the Bell condition (\ref{10}) can be fulfilled. Consider the
spectral decomposition $B^{(\pm )}=\sum_{i}\lambda _{i}E_{B}(\lambda _{i}),$
of an observable $B^{(\pm )}$ with eigenvalues $\lambda _{i}\in \lbrack
-1,1] $ and spectral projections $E_{B^{(\pm )}}(\lambda _{i}).$
Substituting this decomposition into (\ref{10}), we rewrite the Bell
condition (\ref{10}) on perfect correlations/anticorrelations in the form 
\begin{equation}
\sum_{i,k}(\lambda _{i}\lambda _{k}\mp 1)\text{ }\mathrm{tr}[\rho _{d\times
d}\{E_{B^{(\pm )}}(\lambda _{i})\otimes E_{B^{(\pm )}}(\lambda _{k})\}]=0,
\label{12}
\end{equation}%
where all joint probabilities $\mathrm{tr}[\rho _{d\times d}\{E_{B^{(\pm
)}}(\lambda _{i})\otimes E_{B^{(\pm )}}(\lambda _{k})\}]\geq 0$ and sum up
to $1.$This form of condition (\ref{10}) implies that, for an observable $%
B^{(\pm )}$ with eigenvalues $\left\vert \lambda _{i}\right\vert \leq 1,$
condition (\ref{10}) is fulfilled \emph{if and only if }the operator norm $%
\left\Vert B^{(\pm )}\right\Vert =1$ and, for eigenvalues $\lambda _{i},$ $%
\lambda _{k}$, for which the product $\lambda _{i}\lambda _{k}\neq \pm 1,$
the corresponding joint probability $\mathrm{tr}[\rho _{d\times
d}\{E_{B^{(\pm )}}(\lambda _{i})\otimes E_{B^{(\pm )}}(\lambda _{k})\}]=0.$
\end{remark}

For $d=2,$ we specified in Proposition 2 of \cite{KL} a necessary and
sufficient condition for a symmetric two-qubit state $\rho _{2\times 2}$ to
exhibit perfect correlations/anticorrelations if some qubit spin observable $%
\sigma _{b}$ is measured at both sites. We proved (see Theorem 1 in \cite{KL}%
) that, for each two-qubit state $\rho _{2\times 2}$ with perfect
correlations/anticorrelations, the maximal value of the left-hand side (\ref%
{9}) of the original Bell inequality (\ref{7}) over all qubit spin
observables $\sigma _{a},\sigma _{b},\sigma _{\widetilde{b}}$ cannot exceed $%
\frac{3}{2}$ and specified two-qubit states, where this upper bound is
attained: 
\begin{equation}
\max_{\rho _{2\times 2},\sigma _{a},\sigma _{b^{(\pm )}},\sigma _{\widetilde{%
b}}}\text{ }\mathcal{B}_{\rho _{2\times 2}}^{ob}(\sigma _{a},\sigma
_{b^{(\pm )}},\sigma _{\widetilde{b}}){\LARGE |}_{\text{perfect}}=\frac{3}{2}%
.  \label{14}
\end{equation}

In Theorem 2 of \cite{KL} we also showed that, for each symmetric two-qutrit
state $\rho _{3\times 3}$ exhibiting perfect correlations/anticorrelations
if some qutrit spin observable $S_{b}$ is measured at both sites, the
maximal value of the left- hand side (\ref{9}) of the original Bell
inequality over qutrit spin observables $S_{a},S_{b},S_{\widetilde{b}}$ also
admits the bound 
\begin{equation}
\max_{\rho _{3\times 3},\text{ }S_{a},S_{b^{(\pm )}},S_{\widetilde{b}}}\text{
}\mathcal{B}_{\rho _{3\times 3}}^{ob}(S_{a},S_{b},S_{\widetilde{b}}){\LARGE |%
}_{\text{perfect}}\leq \frac{3}{2}.  \label{15}
\end{equation}%
We stressed in \cite{KL} that, in (\ref{14}), (\ref{15}), the attained upper
bound $\frac{3}{2}$ on the maximal violation of the original Bell inequality
is less than the upper bound $\left( 2\sqrt{2}-1\right) $ for the value of $%
\mathcal{B}_{\rho _{d\times d}}(A,B^{(\pm )},\widetilde{B}){\LARGE |}_{\text{%
perfect}}$ following from the Tsirelson upper bound for the CHSH inequality.

In the following section, we proceed to analyse the maximal value of the
left-hand side (\ref{9}) of the original Bell inequality in a general
two-qudit case ($d\geq 2)$.

\section{Quantum violation in a general two-qudit case}

Under the Bell condition (\ref{10}), let us analyze the maximal value of the
Bell expression (\ref{9}) over traceless qudit observables $A,$ $B,$ $%
\widetilde{B}$ with eigenvalues in $[-1,1].$ For short, we further denote
this set of observables by $\mathcal{L}_{d}$.

For an observable $X\in \mathcal{L}_{d}$, consider the normalized version 
\cite{L3} 
\begin{eqnarray}
X &=&\sqrt{\frac{d}{2}}\left( r\cdot \Lambda \right) ,\text{ \ }r_{j}=\frac{1%
}{\sqrt{2d}}\mathrm{tr}[X\Lambda _{j}],  \label{16} \\
\mathrm{tr}[X^{2}] &=&d\left\Vert r\right\Vert ^{2},\text{ \ \ }r\in \mathbb{%
R}^{d^{2}-1},  \notag
\end{eqnarray}%
of the generalized Gell-Mann representation\footnote{%
On the generalized Gell-Mann representation for qudit states see \cite{10,
11, 12} and references therein.}. Here, notation $\left\Vert \cdot
\right\Vert $ means the Euclidian norm of a vector $n$ in $\mathbb{R}%
^{d^{2}-1}$ and $\Lambda _{j},$ $j=1,....,d^{2}-1$ \ are traceless Hermitian
operators on $\mathbb{C}^{d}$ (generators of SU$(d)$ group), satisfying the
relation $\mathrm{tr}[\Lambda _{j}\Lambda _{j_{1}}]=2\delta _{jj_{1}},$ and
presented in Appendix A. The matrix representations of $\Lambda _{j},$ $%
j=1,....,d^{2}-1$ constitute the higher dimensional extensions of the Pauli
matrices for qubits ($d=2$) and the Gell-Mann matrices for qutrits ($d=3$).

As proved in \cite{L3}, representation (\ref{16}) establishes the one-to-one
correspondence 
\begin{equation}
\mathcal{L}_{d}\leftrightarrow \mathfrak{R}_{d}  \label{17}
\end{equation}%
between traceless observables in $\mathcal{L}_{d}$ and $(d^{2}-1)$%
-dimensional vectors $r\in \mathbb{R}^{d^{2}-1}$ in the set 
\begin{equation}
\mathfrak{R}_{d}:=\left\{ r\in \mathbb{R}^{d^{2}-1}\mid \left\Vert r\cdot
\Lambda \right\Vert _{0}\leq \sqrt{\frac{2}{d}}\right\}  \label{18}
\end{equation}%
which is a subset 
\begin{eqnarray}
\mathfrak{R}_{d} &\subseteq &\left\{ r\in \mathbb{R}^{d^{2}-1}\mid
\left\Vert r\right\Vert \leq l_{d}\right\} ,  \label{19} \\
l_{d} &=&1,\ \ \text{if }d\geq 2\text{ \ is even, \ \ \ }l_{d}=\sqrt{\frac{%
d-1}{d}},\text{ \ \ if }d\geq 2\text{ \ is odd,}  \notag
\end{eqnarray}%
of the ball of radius $l_{d}$ in $\mathbb{R}^{d^{2}-1}$ and also contains
the ball $\mathfrak{R}_{d}\supseteq \left\{ n\in \mathbb{R}^{d^{2}-1}\mid
\left\Vert n\right\Vert =\sqrt{\frac{1}{d-1}}\right\} .$ In (\ref{18}),
notation $\left\Vert \cdot \right\Vert _{0}$ means the operator norm of
observables on $\mathbb{C}^{d}.$

If a qudit dimension $d\geq 2$ is even, then (\ref{16}) establishes \cite{L3}
the one-to-one correspondence%
\begin{equation}
\mathcal{L}_{d}\supset \mathcal{L}_{d}^{(0)}\leftrightarrow \mathfrak{R}%
_{d}^{(0)}\subset \mathfrak{R}_{d}  \label{20}
\end{equation}%
between traceless qudit observables with eigenvalues $\pm 1$ (i. e in subset 
$\mathcal{L}_{d}^{(0)}\subset \mathcal{L}_{d}$) and vectors in the
intersection $\mathfrak{R}_{d}^{(0)}$ of $\mathfrak{R}_{d}$ with the unit
sphere: 
\begin{equation}
\mathfrak{R}_{d}^{(0)}=\left\{ r\in \mathbb{R}^{d^{2}-1}\mid \left\Vert
r\cdot \Lambda \right\Vert _{0}=\sqrt{\frac{2}{d}},\text{ \ \ }\left\Vert
r\right\Vert =1\right\} .  \label{21}
\end{equation}%
For details, see Proposition 1 in Section 2 of \cite{L3}.

For each of three qudit observables $A,B,\widetilde{B}\in \mathcal{L}_{d}$
standing in expression (\ref{9}), we specify representation (\ref{16}) as%
\begin{eqnarray}
A &=&\sqrt{\frac{d}{2}}\left( a\cdot \Lambda \right) ,\text{ \ \ }B^{(\pm )}=%
\sqrt{\frac{d}{2}}\left( b^{(\pm )}\cdot \Lambda \right) ,\text{\ \ \ }%
\widetilde{B}=\sqrt{\frac{d}{2}}b\cdot \Lambda ,  \label{22} \\
a,b^{(\pm )},\widetilde{b} &\in &\mathfrak{R}_{d}.  \notag
\end{eqnarray}%
This implies the following expression for the quantum expectation (\ref{8}):%
\begin{equation}
\mathrm{tr}[\rho _{d\times d}\{A\otimes B\}]=\left\langle a,T_{\rho
_{d\times d}}b\right\rangle :=\sum_{n,m}T_{\rho _{d\times
d}}^{(n,m)}a_{n}b_{m},  \label{23}
\end{equation}%
where $T_{\rho _{d\times d}}$ is the linear operator on $\mathbb{R}%
^{d^{2}-1},$ defined in the canonical basis of $\mathbb{R}^{d^{2}-1}$ by the 
$\left( d^{2}-1\right) \times \left( d^{2}-1\right) $ correlation matrix 
\begin{equation}
T_{\rho _{d\times d}}^{(nm)}:=\mathrm{tr}[\rho _{d\times d}\{\Lambda
_{n}\otimes \Lambda _{m}\}],\text{ \ \ }n,m=1,...,d^{2}-1,  \label{24}
\end{equation}%
introduced for an arbitrary qudit dimension $d\geq 2$ in \cite{L3}. This
matrix constitutes a generalization to higher dimensions of the two-qubit
correlation matrix considered in \cite{KL, 13}. For a symmetric\footnote{%
In the sense that $\rho _{d\times d}$ is invariant under permutation of
spaces $\mathbb{C}^{d}$ in the tensor product $\mathbb{C}^{d}\otimes \mathbb{%
C}^{d}.$} two-qudit state $\rho _{d\times d},$ the operator $T_{\rho
_{d\times d}}$ on $\mathbb{R}^{d^{2}-1}$ and its matrix representation (\ref%
{24}) are hermitian.

Substituting (\ref{23}) into relation (\ref{9}) and condition (\ref{10}),
we, correspondingly, derive:%
\begin{eqnarray}
\mathcal{B}_{\rho _{d\times d}}^{ob}(A,B^{(\pm )},\widetilde{B})
&=&\left\vert \text{ }\mathrm{tr}[\rho _{d\times d}\{A\otimes B^{(\pm )}\}]-%
\mathrm{tr}[\rho _{d\times d}\{A\otimes \widetilde{B}\}]\text{ }\right\vert
\pm \mathrm{tr}[\rho _{d\times d}\{B^{(\pm )}\otimes \widetilde{B}\}]  \notag
\\
&=&\frac{d}{2}\left( \left\vert \left\langle a,T_{\rho _{d\times d}}(b^{(\pm
)}-\widetilde{b})\right\rangle \right\vert \pm \left\langle b^{(\pm
)},T_{\rho _{d\times d}}\widetilde{b}\right\rangle \right)  \label{25} \\
&&  \notag \\
\left\langle b^{(\pm )},T_{\rho _{d\times d}}b^{(\pm )}\right\rangle &=&\pm 
\frac{2}{d},\text{ \ \ \ \ }a,b^{(\pm )},\widetilde{b}\in \mathfrak{R}_{d}.
\label{26}
\end{eqnarray}

In what follows, we consider only symmetric two-qudit states $\rho _{d\times
d}$ -- in this case, the correlation matrix $\left( T_{\rho _{d\times
d}}^{(nm)}\right) $ is hermitian, and also, in view of Remark 1 -- only
qudit observables $A,B^{(\pm )},\widetilde{B}$ in $\mathcal{L}_{d}^{(0)},$
that is, traceless and with eigenvalues $\pm 1.$ From relation (\ref{20})
proved in \cite{L3} it follows that, under representation (\ref{16}), these
observables are bijectively mapped to vectors in subset $\mathfrak{R}%
_{d}^{(0)}\subset \mathfrak{R}_{d}$ given by (\ref{21}).

For the maximal value of the Bell expression $\mathcal{B}_{\rho _{d\times
d}}^{ob}(A,B^{(\pm )},\widetilde{B})$ over traceless observables $A,B^{(\pm
)},\widetilde{B}\in \mathcal{L}_{d}^{(0)}$ under constraint (\ref{10}),
relation (\ref{25}) and the one-to-one correspondence (\ref{20}) imply%
\begin{eqnarray}
&&\max_{A,B^{(\pm )},\widetilde{B}\in \mathcal{L}_{d}^{(0)}}\text{ }\mathcal{%
B}_{\rho _{d\times d}}^{ob}(A,B^{(\pm )},\widetilde{B}){\LARGE |}_{\text{%
perfect}}\text{ }  \label{27} \\
&=&\max_{a,b^{(\pm )},\widetilde{b}\in \mathfrak{R}_{d}^{(0)}}\frac{d}{2}%
\left( \left\vert \left\langle a,T_{\rho _{d\times d}}(b^{(\pm )}-\widetilde{%
b})\right\rangle \right\vert \pm \left\langle b^{(\pm )},T_{\rho _{d\times
d}}\widetilde{b}\right\rangle \right) {\LARGE |}_{\text{perfect}}  \notag
\end{eqnarray}%
where, for the maximum standing in the second line, the Bell condition on
perfect correlations/anticorrelations is given by (\ref{26}) -- the plus
sign corresponds to perfect correlations and the minus sign -- to perfect
anticorrelations.

Let us first analyze when a symmetric two-qudit state may satisfy the Bell
condition (\ref{26}), equivalently, (\ref{10}).

\subsection{Two-qudit states with perfect correlations/anticorrelations}

For the hermitian matrix $\left( T_{\rho _{d\times d}}^{nm}\right) ,$ let $%
\lambda _{m}$ be an eigenvalue with a multiplicity $k_{\lambda _{m}}$, $%
\sum_{m}k_{\lambda _{m}}=d^{2}-1,$ and $\mathrm{v}_{\lambda _{m}}^{(j)}\in 
\mathbb{R}^{d^{2}-1},j=1,...,k_{\lambda _{m}},$ be mutually orthogonal unit
eigenvectors 
\begin{equation}
T_{\rho _{d\times d}}\mathrm{v}_{\lambda _{m}}^{(j)}=\lambda _{m}\mathrm{v}%
_{\lambda _{m}}^{(j)}  \label{28}
\end{equation}%
corresponding to this $\lambda _{m}$. Note that the spectral norm of the
correlation matrix (\ref{24}) satisfies the relation 
\begin{equation}
\left\Vert T_{\rho _{d\times d}}\right\Vert =\max_{m}\left\vert \lambda
_{m}\right\vert .  \label{29}
\end{equation}%
Decomposing in (\ref{27}) a unit vector $b^{(\pm )}\in \mathfrak{R}%
_{d}^{(0)} $ ($d\geq 2$ is even) via the orthonormal basis $\{\mathrm{v}%
_{\lambda _{m}}^{(j)}\}:$ 
\begin{equation}
b^{(\pm )}=\sum_{m,j}\beta _{jm}^{(\pm )}\mathrm{v}_{\lambda _{m}}^{(j)}\in 
\mathfrak{R}_{d}^{(0)},\text{ \ }\sum_{m,j}\left( \beta _{jm}^{(\pm
)}\right) ^{2}=1,  \label{30}
\end{equation}%
we rewrite the Bell condition (\ref{26}) in the form%
\begin{equation}
\sum_{m,j}\left( \lambda _{m}\mp \frac{2}{d}\right) \left( \beta _{jm}^{(\pm
)}\right) ^{2}=0.  \label{31}
\end{equation}%
This form implies the following statement.

\begin{proposition}[Sufficient condition]
Let, for a symmetric two-qudit state $\rho _{d\times d}$ with an even $d\geq
2,$ the correlation matrix $T_{\rho _{d\times d}}$ have the spectral norm 
\begin{equation}
\left\Vert T_{\rho _{d\times d}}\right\Vert =\frac{2}{d}  \label{32}
\end{equation}%
and, for the eigenvalue $\lambda _{m_{0}}$ of $T_{\rho _{d\times d}}$ with
maximal absolute value $\left\vert \lambda _{m_{_{0}}}\right\vert =\frac{2}{d%
}$, there exist a unit eigenvector $\mathrm{v}_{\lambda _{m_{0}}}$\
belonging to set $\mathfrak{R}_{d}^{(0)}$ given by (\ref{21}) and
satisfying, therefore, the relations:%
\begin{equation}
\left\Vert \mathrm{v}_{\lambda _{m_{0}}}\cdot \Lambda \right\Vert _{0}=\sqrt{%
\frac{2}{d}},\text{ \ \ }\left\Vert \mathrm{v}_{\lambda _{m_{0}}}\right\Vert
=1.  \label{33}
\end{equation}%
Then, for this state $\rho _{d\times d}$, the Bell condition in the form (%
\ref{26}) is fulfilled on each vector 
\begin{equation}
b^{(\pm )}=\mathrm{v}_{\pm \frac{2}{d}}\in \mathfrak{R}_{d}^{(0)}
\label{33'}
\end{equation}%
and, correspondingly, the Bell condition in the form (\ref{10}) -- for each
qudit observable 
\begin{equation}
B^{(\pm )}=\sqrt{\frac{2}{d}}\left( \mathrm{v}_{\pm \frac{2}{d}}\cdot
\Lambda \right) \in \mathcal{L}_{d}^{(0)}.  \label{34}
\end{equation}%
If $\lambda _{m_{_{0}}}=\frac{2}{d},$ then the Bell condition on perfect
correlations (plus sign) is fulfilled and if $\lambda _{m_{_{0}}}=-\frac{2}{d%
}$, then Bell condition on perfect anticorrelations\ (minus sign) is
fulfilled.
\end{proposition}

In view of this statement, in what follows, we use the following terminology.

\begin{definition}
Denote by $\mathfrak{S}_{d\times d}^{(sym)}$ the class of symmetric
two-qudit states satisfying relations (\ref{32}), (\ref{33}) and, therefore,
exhibiting perfect correlations/anticorrelations if at least one of qudit
observables in $\mathcal{L}_{d}^{(0)}$ is measured at both sites.
\end{definition}

In a two-qubit case, condition (\ref{32}) reduces to $\left\Vert T_{\rho
_{2\times 2}}\right\Vert =1$ while relation (\ref{33}) is fulfilled for each
unit eigenvector of $T_{\rho _{2\times 2}}$. Therefore, in a two-qubit case,
Proposition 1 of the present article reduces to our Proposition 2 in \cite%
{KL}.

Moreover, since, in a two-qubit case relation $\left\Vert T_{\rho _{2\times
2}}\right\Vert \leq 1$ holds for each two-qubit state, condition (\ref{32})
becomes \emph{necessary and sufficient} for a two-qubit state to exhibit
perfect correlations.

As an example of higher dimensional two-qudit states belonging to class $%
\mathfrak{S}_{d\times d}^{(sym)},$ consider the two-qudit
Greenberger--Horne--Zeilinger (GHZ) state 
\begin{equation}
\rho _{ghz,d}=\frac{1}{d}\sum\limits_{j,k=1,...d}|j\rangle \left\langle
k\right\vert \otimes |j\rangle \left\langle k\right\vert .  \label{35}
\end{equation}%
For $d=2$, this state constitutes one of Bell states and its correlation
matrix has the form \cite{KL} 
\begin{equation}
\begin{pmatrix}
1 & 0 & 0 \\ 
0 & -1 & 0 \\ 
0 & 0 & 1%
\end{pmatrix}
\label{36}
\end{equation}%
Therefore, $\left\Vert T_{\rho _{ghz,2}}\right\Vert =1$. Also, as mentioned
above, for $d=2,$ relation (\ref{33}) is fulfilled for all unit eigenvectors
of $T_{\rho _{ghz,2}}.$ Therefore, the two-qubit GHZ state belongs to the
class $\mathfrak{S}_{2\times 2}^{(sym)}.$

Consider an even $d>2.$ As it is proved in \cite{L3}, for the GHZ state $%
\rho _{ghz,d}$ with an arbitrary $d\geq 2$ (not necessarily even), the
correlation matrix $T_{\rho _{ghz,d}}$ has the block diagonal form%
\begin{equation}
\begin{pmatrix}
T^{(s)} & 0 & 0 \\ 
0 & T^{(as)} & 0 \\ 
0 & 0 & T^{(d)}%
\end{pmatrix}
\label{37}
\end{equation}%
where (i) $T^{(s)}$ is the $\frac{d\left( d-1\right) }{2}\times \frac{%
d\left( d-1\right) }{2}$ diagonal matrix with all eigenvalues equal to $%
\frac{2}{d}$; (ii) $T^{(as)}$ is the $\frac{d\left( d-1\right) }{2}\times 
\frac{d\left( d-1\right) }{2}$ diagonal matrix with all eigenvalues equal to 
$(-\frac{2}{d})$; and (iii) $T^{(d)}$ is the $(d-1)\times (d-1)$ diagonal
matrix with eigenvalues $\frac{2}{d}$.

Therefore, for the two-qudit GHZ\ state, the spectral norm of its
correlation matrix $T_{\rho _{ghz,d}}$ is equal to 
\begin{equation}
\left\Vert T_{\rho _{ghz,d}}\right\Vert =\frac{2}{d},\text{ \ \ }\forall
d\geq 2,  \label{38}
\end{equation}%
so that condition (\ref{32}) of Proposition 1 is fulfilled.

From (\ref{37}) it also follows that, for the GHZ state (\ref{35}), the
hermitian matrix $T_{\rho _{ghz,d}}$ has two eigenvalues $\pm \frac{2}{d}$,
hence, two proper subspaces $\mathfrak{J}_{\pm \frac{2}{d}}\subset \mathbb{R}%
^{d^{2}-1}$, so that each vector $r\in \mathfrak{R}_{d}^{(0)}$ is decomposed
as $r=r^{(+)}+r^{(-)},$ where $r^{(\pm )}$ are projections of $r\in 
\mathfrak{R}_{d}^{(0)}$ on the proper subspaces $\mathfrak{J}_{\pm \frac{2}{d%
}}$ and constitute eigenvectors (in general, not unit) of $T_{\rho _{ghz,d}}$
corresponding to eigenvalues $(\pm \frac{2}{d}):$ 
\begin{equation}
T_{\rho _{ghz,d}}r^{(\pm )}=\pm \frac{2}{d}r^{(\pm )},\text{ \ }\left\langle
r^{(+)},r^{(-)}\right\rangle =0,\text{\ }  \label{39}
\end{equation}

Let a dimension $d\geq 2$ be even, hence, subset $\mathcal{L}_{d}^{(0)}$ of
traceless qudit observables with eigenvalues $\pm 1$ be not empty. Consider
a qudit observable $X\in \mathcal{L}_{d}^{(0)}$ of the form%
\begin{equation}
X=\tsum\limits_{m=1,2,..,d}(-1)^{\gamma _{m}}|m\rangle \langle m|\text{, \ \
\ }\tsum\limits_{m}(-1)^{\gamma _{m}}=0,\text{ \ \ \ \ even }d\geq 2,
\label{40}
\end{equation}%
where (i) $\gamma _{m}$ are arbitrarily chosen positive integers $\gamma
_{m} $ guaranteeing \textrm{tr}$[X]=0$; (ii) $|m\rangle $ are mutually
orthogonal unit vectors of the computational basis $\{|m\rangle ,$ $%
m=1,....,d\}$ in $\mathbb{C}^{d}.$

Due to the structure (\ref{A1}) of operators $\Lambda _{jk}^{(s)}$, $\Lambda
_{jk}^{(as)}$, $\Lambda _{l}^{(d)}$, under representation (\ref{16}), to an
observable of the form (\ref{40}) there corresponds the unit vector $%
r_{X}\in \mathfrak{R}_{d}^{(0)},$ for which the projection $r_{X}^{(-)}=0,$
therefore, $r_{X}=r_{X}^{(+)}\in $ $\mathfrak{R}_{d}^{(0)}\cap \mathfrak{J}_{%
\frac{2}{d}}.$

Also, if we take a qudit observable $X^{\prime }\in \mathcal{L}_{d}^{(0)}$
of the form%
\begin{eqnarray}
X^{\prime } &=&\tsum\limits_{m=1,3,5,...,d-1}(-1)^{\gamma _{m}}\left( \frac{%
(|m+1\rangle +|m\rangle )(\langle m+1|\text{ }+\left\langle m\right\vert )}{2%
}-\frac{(|m+1\rangle -|m\rangle )(\langle m+1|\text{ }-\left\langle
m\right\vert )}{2}\right)  \notag \\
&=&\tsum\limits_{m=1,3,5,..,d-1}(-1)^{\gamma _{m}}\left( |m+1\rangle
\left\langle m\right\vert +|m\rangle \langle m+1|\right) ,\text{ \ \ \ even }%
d\geq 2,  \label{41}
\end{eqnarray}%
with arbitrarily chosen positive integers $\gamma _{m},$ then, under
representation (\ref{16}), to an observable (\ref{41}) there also
corresponds the unit vector $r_{X^{\prime }}\in \mathfrak{R}_{d}^{(0)},$ for
which the projection $r_{X^{\prime }}^{(-)}=0,$ so that $r_{X^{\prime
}}=r_{X^{\prime }}^{(+)}\in $ $\mathfrak{R}_{d}^{(0)}\cap \mathfrak{J}_{%
\frac{2}{d}}.$

However, to a qudit observable $X^{\prime \prime }\in \mathcal{L}_{d}^{(0)}$
of the form 
\begin{eqnarray}
X^{\prime \prime } &=&\tsum\limits_{m=1,3,5,...,d-1}(-1)^{\gamma _{m}}\left( 
\frac{(|m+1\rangle +i|m\rangle )(\langle m+1|\text{ }+i\left\langle
m\right\vert )}{2}-\frac{(|m+1\rangle -i|m\rangle )(\langle m+1|\text{ }%
-i\left\langle m\right\vert )}{2}\right)  \notag \\
&=&\tsum\limits_{m=1,3,5,...,d-1}(-1)^{\gamma _{m}}\left( -i\text{ }%
|m+1\rangle \left\langle m\right\vert +i\text{ }|m\rangle \langle
m+1|\right) ,\text{ \ \ \ even }d\geq 2,  \label{42}
\end{eqnarray}%
with arbitrary positive integers $\gamma _{m},$ under representation (\ref%
{16}), there corresponds the unit vector $r_{X^{\prime \prime }}\in 
\mathfrak{R}_{d}^{(0)}$ with projection $r_{X^{\prime \prime }}^{(+)}=0,$ so
that this observable is mapped to the unit eigenvector $r_{X^{\prime \prime
}}=r_{X}^{(-)}\in $ $\mathfrak{R}_{d}^{(0)}\cap \mathfrak{J}_{(-\frac{2}{d}%
)} $ of $T_{\rho _{ghz,d}}$ corresponding to eigenvalue $(-\frac{2}{d}).$

Thus: (i) from (\ref{38}) it follows that, for the GHZ state (\ref{35}),
condition (\ref{32}) is fulfilled; (ii) from (\ref{40})--(\ref{42}) and (\ref%
{20}), (\ref{21}) it follows that, in case of GHZ state $\rho _{ghz,d}$ with
an even $d\geq 2,$ for eigenvalues $\pm \frac{2}{d}$ of the correlation
matrix $T_{\rho _{ghz,d}}$, there exist unit eigenvectors which belong to
set $\mathfrak{R}_{d}^{(0)}.$

Therefore, by Proposition 1, the GHZ state $\rho _{ghz,d}$ exhibits perfect
correlations if any of observables $X,X^{\prime }\in \mathcal{L}_{d}^{(0)}$
of forms (\ref{40}), (\ref{41}) is measured at both sites and perfect
anticorrelations if at both sites any observable $X^{\prime \prime }\in 
\mathcal{L}_{d}^{(0)}$ of form (\ref{42}) is measured.

Summing up, we have proved the following statement.

\begin{proposition}
For an arbitrary even $d\geq 2,$ the two-qudit GHZ state (\ref{35}) belongs
to the class $\mathfrak{S}_{d\times d}^{(sym)}$ of two-qudit states
exhibiting perfect correlations/anticorrelations.
\end{proposition}

\subsection{ Quantum analog}

For a state $\rho _{d\times d}\in \mathfrak{S}_{d\times d}^{(sym)}$ with an
even dimension $d\geq 2$ and a vector $b^{(\pm )}\in \mathfrak{R}_{d}^{(0)}$
specified in Proposition 1, let us now analyze the value of the maximum (\ref%
{27}).

Since by (\ref{21}) $\mathfrak{R}_{d}^{(0)}$ is a subset of the unit sphere
in $\mathbb{R}^{d^{2}-1},$ we have the following relation:%
\begin{eqnarray}
&&\frac{d}{2}\max_{a,\widetilde{b}\in \mathfrak{R}_{d}^{(0)}}\text{ }\left(
\left\vert \left\langle a,T_{\rho _{d\times d}}(b^{(\pm )}-\widetilde{b}%
)\right\rangle \right\vert \pm \left\langle b^{(\pm )},T_{\rho _{d\times d}}%
\widetilde{b}\right\rangle \right) {\LARGE |}_{\text{perfect}}  \label{43} \\
&\leq &\frac{d}{2}\max_{\left\Vert a\right\Vert ,\left\Vert \widetilde{b}%
\right\Vert =1}\text{ }\left( \left\vert \left\langle a,T_{\rho _{d\times
d}}(b^{(\pm )}-\widetilde{b})\right\rangle \right\vert \pm \left\langle
b^{(\pm )},T_{\rho _{d\times d}}\widetilde{b}\right\rangle \right) {\LARGE |}%
_{\text{perfect}}.  \notag
\end{eqnarray}%
In (\ref{43}), the maximum of the expression 
\begin{equation}
\frac{d}{2}\left( \left\vert \left\langle a,T_{\rho _{d\times d}}(b^{(\pm )}-%
\widetilde{b})\right\rangle \right\vert \pm \left\langle b^{(\pm )},T_{\rho
_{d\times d}}\widetilde{b}\right\rangle \right)  \label{44}
\end{equation}%
over unit vectors $\left\Vert a\right\Vert =1$ is attained on the unit vector%
\begin{equation}
\widetilde{a}=\frac{T_{\rho _{d\times d}}(b^{(\pm )}-\widetilde{b})}{%
\left\Vert T_{\rho _{d\times d}}(b^{(\pm )}-\widetilde{b})\right\Vert }
\label{45}
\end{equation}%
and is equal to 
\begin{equation}
\frac{d}{2}\left( \left\Vert T_{\rho _{d\times d}}(b^{(\pm )}-\widetilde{b}%
)\right\Vert \pm \left\langle b^{(\pm )},T_{\rho _{d\times d}}\widetilde{b}%
\right\rangle \right) .  \label{46}
\end{equation}

For a vector $b^{(\pm )}\in \mathfrak{R}_{d}^{(0)}$, specified in
Proposition 1, decomposition (\ref{30}) reads 
\begin{equation}
b^{(\pm )}=\sum_{j=1,...,k_{\lambda _{_{m_{_{0}}}}}}\beta _{jm_{_{0}}}^{(\pm
)}\mathrm{v}_{\lambda _{m_{_{0}}}}^{(j)},\text{ \ }\sum_{j=1,...,k_{\lambda
_{_{m_{_{0}}}}}}\left( \beta _{jm_{_{0}}}^{(\pm )}\right) ^{2}=1.  \label{47}
\end{equation}%
Expanding also a vector $\widetilde{b}=\sum_{m,j}\widetilde{\beta }_{jm}%
\mathrm{v}_{\lambda _{m}}^{(j)}$, $\sum_{m,j}\left( \widetilde{\beta }%
\right) _{jm}^{2}=1$ via the orthonormal basis $\left\{ \mathrm{v}_{\lambda
_{m}}^{(j)}\right\} $ and substituting this decomposition and decomposition (%
\ref{47}) into (\ref{46}), we derive%
\begin{eqnarray}
&&\text{ }\frac{d}{2}\left( \left\Vert T_{\rho _{d\times d}}(b^{(\pm )}-%
\widetilde{b})\right\Vert \pm \left\langle b^{(\pm )},T_{\rho _{d\times d}}%
\widetilde{b}\right\rangle \right) {\LARGE |}_{\text{perfect}}  \label{48} \\
&&  \notag \\
&=&\frac{d}{2}\left( \left\vert \sqrt{\sum \lambda _{m}^{2}\left( \beta
_{mj}^{(\pm )}-\widetilde{\beta }_{mj}\right) ^{2}}\right\vert \pm \sum
\lambda _{m}\beta _{mj}^{(\pm )}\widetilde{\beta }_{mj}\right) {\LARGE |}_{_{%
\text{{\large perfect}}}}  \notag \\
&&  \notag \\
&=&\text{ }\sqrt{2\left( 1-\sum_{j=1}^{k_{\lambda _{m_{_{0}}}}}\beta
_{jm_{_{0}}}^{(\pm )}\widetilde{\beta }_{jm_{_{0}}}\right) -\frac{d^{2}}{4}%
\sum_{m,j}(\frac{4}{d^{2}}-\lambda _{m}^{2})\left( \widetilde{\beta }%
_{jm}\right) ^{2}}+\sum_{j=1}^{k_{\lambda _{m_{_{0}}}}}\beta
_{jm_{_{0}}}^{(\pm )}\widetilde{\beta }_{jm_{_{0}}}  \notag
\end{eqnarray}%
Since, for a state $\rho _{d\times d}\in \mathfrak{S}_{d\times d}^{(sym)}$,
all eigenvalues $\lambda _{m}^{2}\leq \frac{4}{d^{2}},$ from the expression
in the last line of (\ref{48}) it follows that, for all choices of a vector $%
b^{(\pm )}\in \mathfrak{R}_{d}^{(0)}$, specified for a state $\rho _{d\times
d}\in \mathfrak{S}_{d\times d}^{(sym)}$ in Proposition 1, maximum (\ref{43})
admits the bound%
\begin{eqnarray}
&&\max_{a,\widetilde{b}\in \mathfrak{R}_{d}^{(0)}}\text{ }\frac{d}{2}\left(
\left\vert \left\langle a,T_{\rho _{d\times d}}(b^{(\pm )}-\widetilde{b}%
)\right\rangle \right\vert \pm \left\langle b^{(\pm )},T_{\rho _{d\times d}}%
\widetilde{b}\right\rangle \right) \text{{\LARGE \TEXTsymbol{\vert}}}_{\text{%
perfect}}  \label{49} \\
&\leq &\max_{z\in \lbrack -1,1]}\left( \sqrt{2(1-z)}+z\right) =\frac{3}{2}. 
\notag
\end{eqnarray}

This implies that, for each state $\rho _{d\times d}\in \mathfrak{S}%
_{d\times d}^{(sym)}$ with an even $d\geq 2,$ the maximal value of the
left-hand side of the original Bell inequality over all traceless qudit
observables with eigenvalues $\pm 1$ admits the bound 
\begin{eqnarray}
&&\max_{a,b,\widetilde{b}\in \mathfrak{R}_{d}^{(0)}}\text{ }\frac{d}{2}%
\left( \left\vert \left\langle a,T_{\rho _{d\times d}}(b^{(\pm )}-\widetilde{%
b})\right\rangle \right\vert \pm \left\langle b^{(\pm )},T_{\rho _{d\times
d}}\widetilde{b}\right\rangle \right) {\LARGE |}_{\text{perfect}}  \label{50}
\\
&\leq &\frac{3}{2}.  \notag
\end{eqnarray}%
As we have proved by in \cite{KL} (Theorem 1), for $d=2,$ this upper bound
is attained.

Relations (\ref{27}), (\ref{43}) and (\ref{50}) imply the following
statement.

\begin{theorem}
Let a qudit dimension $d\geq 2$ be even and a symmetric two-qudit state $%
\rho _{d\times d}$ belong to the class $\mathfrak{S}_{d\times d}^{(sym)}$ of
states specified in Proposition 1 and exhibiting perfect
correlations/anticorrelations (\ref{10}) whenever a qudit observable $%
B^{(\pm )}$ is measured at both sites. Then the maximal value of the
left-hand side $\mathcal{B}_{\rho _{d\times d}}(A,B^{(\pm )},\widetilde{B})$
of the original Bell inequality (\ref{9})\ over all traceless qudit
observables $A,B^{(\pm )},\widetilde{B}$ with eigenvalues $\pm 1$ admits the
bound 
\begin{equation}
\max_{A,B^{(\pm )},\widetilde{B}\in \mathcal{L}_{d}^{(0)}}\text{ }\mathcal{B}%
_{\rho _{d\times d}}^{ob}(A,B^{(\pm )},\widetilde{B}){\LARGE |}_{\text{%
perfect}}\leq \frac{3}{2}  \label{51}
\end{equation}%
and this upper bound is, for example, attained on two-qubit states specified
in \cite{KL}.
\end{theorem}

We stress that, in (\ref{51}), the general upper bound $\frac{3}{2}$ on the
maximal violation of the original Bell inequality is less than the upper
bound $\left( 2\sqrt{2}-1\right) ,$ which follows for the value of $\mathcal{%
B}_{\rho _{d\times d}}^{ob}(A,B^{(\pm )},\widetilde{B}){\LARGE |}_{\text{%
perfect}}$ from the Tsirelson upper bound for the CHSH inequality.

Theorem 1 proves that, for an arbitrary even $d\geq 2$ and all two-qudit
states $\rho _{d\times d}\in \mathfrak{S}_{d\times d}^{(sym)}$ with perfect
correlations/anticorrelations $\mathrm{tr}[\rho _{d\times d}\{B^{(\pm
)}\otimes B^{(\pm )}\}]=\pm 1,$ \emph{the quantum analog} of the original
Bell inequality 
\begin{equation}
\left( \left\vert \text{ }\mathrm{tr}[\rho \{A\otimes B^{(\pm )}\}]-\mathrm{%
tr}[\rho \{A\otimes \widetilde{B}\}]\text{ }\right\vert \pm \mathrm{tr}[\rho
\{B^{(\pm )}\otimes \widetilde{B}\}]\right) {\LARGE |}_{\text{perfect}}\leq 
\frac{3}{2}  \label{52}
\end{equation}%
holds for all traceless qudit observables $A,B^{(\pm )},\widetilde{B}$ with
eigenvalues $\pm 1.$

\section{Conclusions}

In our recent article \cite{KL}, we introduced \emph{a} \emph{necessary and
sufficient condition }for a symmetric two-qubit state to exhibit perfect
correlations/ anticorrelations and proved\ \cite{KL} that, for all symmetric
two-qubit states exhibiting perfect correlations/ anticorrelations,
violation of the original Bell inequality is bounded by $3/2$ from above and
that this upper bound is attained.

In the present paper, for an even qudit dimension $d\geq 2,$ we have
specified (Proposition 1, Definition 1) a class of two-qudit states
exhibiting perfect correlations/anticorrelations and have proved (Theorem 1)
that, for each two-qudit state from this class, the maximal violation of the
original Bell inequality over all traceless qudit observables with
eigenvalues $\pm 1$ is also bounded from above by the value $3/2$. We have
shown (Proposition 2) that the two-qudit GHZ state with an arbitrary even $%
d\geq 2$ belongs to the introduced two-qudit state class.

These our new results are important steps towards proving the $3/2$ upper
bound\ conjecture on the original Bell inequality violation for all
bipartite quantum states with perfect correlations/anticorrelations. As we
see, the proof of this upper bound for arbitrary higher dimensions is
nontrivial and stimulated our application of the generalized Gell-Mann
representation.

Although the authors expect that the same technique can lead to the proof of
the $3/2$ upper bound in\ a general case, one cannot exclude that the
complexity of calculations would lead to finding other approaches (see,
e.g., \cite{14}). Of course, one still cannot exclude that, for the maximal
violation of the original Bell inequality by a general bipartite state with
perfect correlations/anticorrelations, the upper bound $3/2$\ may not be
true.

The states specified in this article can also be used in experiments to test
violation of the original Bell inequality. For the moment, we cannot guess
whether it would be easier to combine high fidelity and detection efficiency
for such class of states. But one cannot exclude that these states can play
the important role in future tests on violation of the original Bell
inequality.

The authors hope that the new results of the present article would attract
the interest of the quantum information community to theoretical analysis of
the original Bell inequality and its possible experimental testing.

\section*{Acknowledgments}

E. R. Loubenets was supported within the Academic Fund Program at the
National Research University Higher School of Economics (HSE) in 2018-2019
(grant N 18-01-0064) and by the Russian Academic Excellence Project "5-100".

\section{Appendix A}

In (\ref{16}), the traceless hermitian operators have the following
indexation form 
\begin{equation}
(\Lambda _{1},...,\Lambda _{d^{2}-1})\rightarrow (\Lambda
_{12}^{(s)},...,\Lambda _{1d}^{(s)},...,\Lambda _{d-1,d}^{(s)},\Lambda
_{12}^{(as)},...,\Lambda _{1d}^{(as)},...,\Lambda _{d-1,d}^{(as)},\Lambda
_{1}^{(d)},...,\Lambda _{d-1}^{(d)}),  \tag{A1}  \label{A1}
\end{equation}%
where \cite{10, 11, 12}

\begin{eqnarray}
\frac{d(d-1)}{2}\text{\ operators}\text{: } &&\Lambda _{mk}^{s}=\left\vert
m\right\rangle \left\langle k\right\vert +\left\vert k\right\rangle
\left\langle m\right\vert ,\text{ \ \ }\Lambda _{mk}^{s}=\Lambda _{km}^{s},%
\text{ \ \ }1\leq m<k\leq d,  \TCItag{A2}  \label{A2} \\
&&  \notag \\
\frac{d(d-1)}{2}\text{\ operators}\text{: } &&\Lambda _{mk}^{as}=-i\text{ }%
\left\vert m\right\rangle \left\langle k\right\vert +i\text{ }\left\vert
k\right\rangle \left\langle m\right\vert ,\text{ \ \ }\Lambda
_{mk}^{as}=-\Lambda _{km}^{as},\text{ \ }1\leq m<k\leq d,  \notag \\
&&  \notag \\
(d-1)\text{\ \ operators}\text{: } &&\Lambda _{l}^{d}=\sqrt{\frac{2}{l(l+1)}}%
\left( \sum\limits_{m=1,...,l}\left\vert m\right\rangle \left\langle
m\right\vert -l\text{ }\left\vert l+1\right\rangle \left\langle
l+1\right\vert \right) ,\text{ \ }1\leq l\leq d-1.  \notag
\end{eqnarray}%
Here, $\{\left\vert m\right\rangle ,$ $m=1,...,d\}$ is the computational
basis of $\mathbb{C}^{d}.$ The matrix representations of $\Lambda _{j},$ $%
j=1,....,d^{2}-1$ constitute the higher--dimensional extensions of the Pauli
matrices for qubits ($d=2$) and the Gell-Mann matrices for qutrits ($d=3$).

\end{document}